\documentclass[12pt]{article}

\title{\bf Dependence of linear polarization of radiation
 in accretion disks on the spin of central black hole}

\author{N.A. Silant'ev\thanks{E-mail: nsilant@bk.ru},
 M.Yu. Piotrovich, Yu.N. Gnedin, T.M. Natsvlishvili\\
 Central Astronomical Observatory at Pulkovo, 196140,
 Saint-Petersburg, Russia.}

\begin{document}

\maketitle

\begin{abstract}
We suppose that linear optical polarization is due to multiple
scattering in optically thick magnetized accretion disk around
central black hole. The polarization degree is very sensitive to
the spin of black hole - for Kerr rotating hole the polarization
is higher than for Schwarzschild non-rotating one if both holes
have the same luminosities and masses. The reason of this effect
is that the radius of the first stable orbit for non-rotating hole
is equal to three gravitational radiuses, and for fast rotating
Kerr hole is approximately 6 times lesser. Magnetic field,
decreasing from first stable orbits, is much larger in the region
of escaping of optical radiation for the case of Schwarzschild
hole than for Kerr one. Large magnetic field gives rise to large
depolarization of radiation due to Faraday rotation effect. This
explains the mentioned result. It seems that the ensemble of
objects with observed polarization mostly consists of Kerr black
holes.

\end{abstract}

\section{Introduction}

The estimation of the black hole spin, i.e. the dimensionless
parameter $a_* = J / J_{max}=Jc/GM^2$, which determines the
angular momentum, is very difficult from observational point of
view. To some extent the observation of luminosity $L_{bol}=
\varepsilon \dot{M} c^2$ of accretion disk allow us qualitatively
estimate this parameter, if we compare this value with theoretical
models predictions. Here $\dot{M}$ is the mass accretion velocity
and $c$ being light velocity. Dimensionless parameter
$\varepsilon$ demonstrates what part of mass transforms to the
radiation energy. For Schwarzschild's black hole ($a_* = 0$) the
theoretical estimation gives $\varepsilon = 0.057$, and for fast
rotating Kerr hole ($a_* \approx 1$) the value $\varepsilon =
0.42$ (See Blandford (1990), Krolik (2007) and Shapiro (2007)).
So, one can qualitatively consider that the bright objects most
probably correspond to the Kerr black holes, and the objects with
weak luminosity correspond to Schwarzschild ones. Evidently, any
other technique to estimate the parameter $a_*$, may be only
qualitatively, will be of interest in this situation.

In this paper we demonstrate that the degree of linear
polarization from magnetized accretion disk strongly depends on
the parameter $a_*$. We suppose that the main part of polarization
in visual range is due to multiple scattering of radiation in
optically thick magnetized accretion disks. In our model we accept
that the radiation sources are located far from the disk's surface
and, consequently, the linear polarization describes by known
Milne's problem for conservative magnetized atmosphere. Note that
for non-magnetized electron atmosphere the maximal degree of
polarization is equal to $p(\mu = 0) = 11.71\%$ and is observed in
direction ${\bf n}$ perpendicular to surface normal ${\bf N}$ (see
Chandrasekhar 1950).

Here and what follow $\mu = \cos{i}$ is cosine of disk's
inclination angle.

For estimations of linear polarization from magnetized disks we
use the approximate analytical formulae (see Silan'ev 2002 and
Silant'ev et al. 2009). The far located disks we observe as a
point sources. Therefore, we use azimuthally averaged values of
polarization. In this case the integral polarization is described
by simple formulae:

\begin{equation}
 p(\mu,{\bf B}) = \frac{p(\mu)}{\left[(1 + C)^4 + 2(1 + C)^2
 (a^2 + b^2) + (a^2 - b^2)^2 \right]^{1/4}};
 \label{eq1}
\end{equation}

\begin{equation}
 \tan{2\chi} = \frac{2 (1 + C) a}{(p(\mu) / p(\mu,{\bf B}))^2
 + (1 + C)^2 + b^2 - a^2}.
 \label{eq2}
\end{equation}

\noindent Here $p(\mu)$ is the degree of polarization for
non-magnetized atmosphere (Chandrasekhar's value), ${\bf B} ={\bf
B}_{\parallel} + {\bf B}_{\perp}$ is magnetic field inside the
disk, the component ${\bf B}_{\parallel}$ is directed parallel to
the disk normal ${\bf N}$, and $B_{\perp} = \sqrt{B_{\rho}^2 +
B_{\varphi}^2}$ is magnetic field perpendicular to ${\bf N}$.
Dimensionless parameters $a$ and $b$ describe the Faraday
depolarization of radiation:

\begin{equation}
 a = 0.8 \lambda^2 B_{\parallel} \mu \equiv \delta_{\parallel}\mu
 ,\,\,\, b = 0.8 \lambda^2 B_{\perp} \sqrt{1 - \mu^2} \equiv
 \delta_{\perp} \sqrt{1 - \mu^2}
 \label{eq3}
\end{equation}

\noindent In all the paper we take the magnetic field in Gauss,
the wavelengths - in micrometers ($\mu$m), and the distances - in
centimeters. Physically the dimensionless parameter $\delta = 0.8
\lambda^2 B$ corresponds to Faraday rotation of polarization plane
along the Thomson optical length $\tau = 2$, if the magnetic field
directed along line of sight {\bf n}. Positional angle $\chi = 0$
corresponds to wave electric field oscillations in the direction
perpendicular to the plane ({\bf nN}), i.e. parallel to disk's
plane.

Parameter $C$ arises in turbulent magnetized plasma (see Silant'ev
2005) and characterizes new effect - an additional extinction of
Stokes parameters $Q$ and $U$ due to incoherent Faraday rotation
of the polarization plane at small scale turbulent curls.

\begin{equation}
 C \approx 0.64 \tau \lambda^4 \langle B'^{2} \rangle f_{B} / 3
 \label{eq4}
\end{equation}

\noindent where $\tau$ is mean Thomson optical depth of a
turbulent curl , $\langle B'^{2} \rangle$ characterizes mean value
of magnetic field fluctuations , $f_B \approx 1$ - dimensionless
parameter, integrally describing correlation of $B'$  taking at
two near places of an atmosphere.

Below we will use the case ${\bf B}_{\perp}=0$. In this case
formulae (1) and (2) are most simple:

\begin{equation}
 p(\mu,B_{\parallel}) = \frac{p(\mu)}{\sqrt{(1+C)^2+a^2}},\,\,\,
 \tan{2\chi} = \frac{a}{1+C}.
 \label{eq5}
\end{equation}

\noindent If we take that the minimum observed polarization is
equal to 0.1\%, then we can observe the linear polarization in
plasma without turbulent magnetic field ($C=0$) in optical range
($\lambda\approx 0.55\mu $m) at $\mu < 0.7$ and magnetic fields in
the emitting region in the interval from zero up $B_{\parallel}\mu
\approx 400 $. This is rather wide range of magnetic field
variations. Thus, for the object NGC4258 authors (see Silant'ev
2009) estimated the magnetic field in the region of optical
emission as $B_{\parallel}\approx 160 $ Gauss.

\section{Magnetic field in accretion disk} 

Usually one uses that near the black hole horizon the strong
magnetic field ${\bf B}_H$ arises, which spreads into accretion
disk by power law, $\sim r^{-n}$. Remember, that the horizon
radius $R_H$ is determined by the formula:

\begin{equation}
 R_H = \frac{1}{2} r_g (1 + \sqrt{1 - a_*^2}),\,\,\,
 r_g = 2 \frac{G M_{BH}}{c^2} = 2.95\times 10^5
 \left(\frac{M_{BH}}{M_{\odot}}\right)[cm]
 \label{eq6}
\end{equation}

\noindent where $r_g$ is the gravitational radius of the central
mass $M_{BH}$, dimensionless parameter $a_*$ determines the spin
of a black hole, $G$ is gravitational constant, $c$ - the light
velocity. For Schwarzschild's black hole ($a_* = 0$) the value
$R_H=r_g$, and for the fast rotating Kerr's black hole with $a_* =
1$ the value $R_H \approx r_g / 2$.

Regular power law decreasing of magnetic field is accepted from
the radius $r_{ms}$ of first from the center stable orbit.
Analytical expression for $r_{ms}$ has the form (see Zhang 2005,
Murphy 2009):

\[
 r_{ms} \equiv q(a_*) r_g = \frac{1}{2} (3 + A_2 - [(3 - A_1)
 (3 + A_1 + 2 A_2)]^{1/2})r_g,
\]
\begin{equation}
 A_1 = 1 + (1 - a_*^2)^{1/3}[(1 + a_*)^{1/3} + (1 - a_*)^{1/3}],
 \,\,\, A_2 = (3 a_*^2 + A_1^2)^{1/2}.
 \label{eq7}
\end{equation}

\noindent According to this formula, Schwarzschild's hole has
$r_{ms} = 3 r_g$, and fast rotating Kerr's hole ($a_*\approx 1$)
has $r_{ms} = r_g / 2$, therewith the decreasing of $r_{ms}$ near
$a_* \approx 1$ is very fast: $r_{ms} \approx r_g (1 +
(4\delta)^{1/3}) / 2$ at $a_* \approx (1 - \delta)$ $(\delta \ll
1)$ (see Chandrasekhar 1983). The values $q(a_*)$ for a number of
$a_*$ are presented in Table 1.

Usually in accretion disks models one considers that at horizon
$R_H$ the magnetic field ${\bf B}_H$ is perpendicular to the
spherical surface, but further it enters to accretion disk
perpendicular to disk's surface, i.e. in disk exists the component
$B_{\parallel}(r)$ (see Wang 2002, 2003; Zhang 2005).

The values $B_{\parallel}(r_{ms})\equiv k B_H$ are slightly
different for holes with different rotation velocity. Thus, for
$a_* = 1$ the factor $k = 0.5$, and for $a_* = 0$ $k = 1 / 3$
takes place (see Wang 2002). So, we can take

\begin{equation}
 B_{\parallel}(r) = B_{\parallel}(r_{ms})
 \left(\frac{r_{ms}}{r}\right)^n \equiv k q^n(a_*)B_H
 \left(\frac{r_g}{r}\right)^n
 \label{eq8}
\end{equation}

\noindent For power index one usually accepts $1 \leq n < 3$ (see
Pariev 2003).

The value $B_H$ , according to Li 2002, Wang 2002, and Ma 2007, is
presented by the expression:

\[
 B_H = \frac{\sqrt{2 k_m c \dot{M}}}{R_H} \approx
 \frac{1.66 \sqrt{k_m}}{1 + \sqrt{1 - a_*^2}}
 \left(\frac{M_{BH}}{M_{\odot}}\right)^{-1} \dot{M}^{1/2} \approx
\]
\begin{equation}
 \approx \sqrt{k_m} \left(\frac{M_{BH}}{M_{\odot}}\right)^{-1/2}
 \left(\frac{L_{bol}}{\varepsilon L_{edd}}\right)^{1/2}
 \frac{10^{8.8}}{1 + \sqrt{1 - a_*^2}}
 \label{eq9}
\end{equation}

\noindent Remember that the Eddington luminosity $L_{edd} = 1.3
\times 10^{38} (M_{BH} / M_{\odot})$, and the bolometric
luminosity of a disk $L_{bol} = \varepsilon \dot{M} c^2$. The
parameter $k_m = P_{magn} / P_{gas}$ relates magnetic pressure
with the gas pressure into disk. At equilibrium usually one takes
$k_m\approx 1$. The parameter $\varepsilon$, as it was mentioned,
characterizes the effectiveness of transition of accretion energy
(the accretion velocity $\dot{M}$) to form of radiation. This
parameter depends on particular mechanisms of transformation and
also on the spin $a_*$. So, for $a_* = 0$ is found  $\varepsilon =
0.057$, and for $a_* =1$ one takes $\varepsilon = 0.42$ (see
Blandford 1990, Krolik 2007, and Shapiro 2007).

\section{Dependence of the polarization degree in continuum on
 spin of black hole} 

Continuum radiation with wavelength $\lambda$ basically is
generated at characteristic distance $R_{\lambda}$ from the disk
center. According to Poindexter et al. 2008, this distance is
determined by expression:

\begin{equation}
 R_{\lambda} = 0.97 \times 10^{10} \lambda^{4/3}
 \left(\frac{M_{BH}}{M_{\odot}}\right)^{2/3}
 \left(\frac{L_{bol}}{\varepsilon L_{edd}}\right)^{1/3} =
 10^{4.27} \lambda^{4/3}
 \left(\frac{M_{BH}}{M_{\odot}}\right)^{1/3} \dot{M}^{1/3}
 \label{eq10}
\end{equation}

\noindent To obtain the magnetic field value
$B_{\parallel}(R_{\lambda})$ from Eq.(8), we have to know the
ratio

\begin{equation}
 \frac{r_g}{R_{\lambda}} = 10^{-4.52}
 \left(\frac{M_{BH}}{M_{\odot}}\right)^{1/3}
 \left(\frac{L_{bol}}{\varepsilon L_{edd}}\right)^{-1/3}
 \lambda^{-4/3} \approx 10^{1.2} \lambda^{-4/3}
 \left(\frac{M_{BH}}{M_{\odot}}\right)^{2/3} \dot{M}^{-1/3}.
 \label{eq11}
\end{equation}

\noindent As a result, we derive for $B_{\parallel}(R_{\lambda})$
the expressions:

\begin{equation}
 B_{\parallel}(R_{\lambda}) =10^{8.8-4.52n}
 \frac{k q^{n}(a_*)\sqrt{k_m}}{1 + \sqrt{1 - a_*^2}}\lambda^{-4n/3}
 \left(\frac{M_{BH}}{M_{\odot}}\right)^{n/3 - 1/2}
 \left(\frac{L_{bol}}{\varepsilon L_{edd}}\right)^{1/2 - n/3},
 \label{eq12}
\end{equation}

\begin{equation}
 B_{\parallel}(R_{\lambda}) =10^{0.22+1.2n}
 \frac{k q^{n}(a_*)\sqrt{k_m}}{1 + \sqrt{1 - a_*^2}}\lambda^{-4n/3}
 \left(\frac{M_{BH}}{M_{\odot}}\right)^{2n/3 - 1}
 \left(\dot{M}\right)^{1/2 - n/3}.
 \label{eq13}
\end{equation}

\noindent The first expression is convenient for calculation of
$B_{\parallel}(R_{\lambda})$, if is known the bolometric disk
luminosity and the black hole mass. The second formula relates
directly $B_{\parallel}(R_{\lambda})$ with the accretion velocity.
Equations (12) and (13) contain the function $f_n(a_*) =
q^{n}(a_*)/(1 + \sqrt{1 - a_*^2})$, which basically determine the
dependence of magnetic field ($B \sim f_n$) in the radiation
region from the black hole spin. The values of this function are
presented in Table 1.

\begin{table}
 \caption[]{Values of $f_n$ as a function of $a_*$.}
 \label{tab1}
 \centering
 \begin{tabular}{|l|c|c|c|c|c|c|c|c|c|c|}
 \hline
 $a_*$                          & 0     & 0.25  & 0.5  & 0.75 & 0.8  & 0.9  & 0.95 & 0.99 & 0.998& 1 \\
 \hline
 {\tiny $1 + \sqrt{1 - a_*^2}$} & 2     & 1.97  & 1.89 & 1.66 & 1.60 & 1.44 & 1.31 & 1.14 & 1.06 & 1 \\
 $q(a_*)$                       & 3     & 2.98  & 2.12 & 1.58 & 1.45 & 1.16 & 0.97 & 0.71 & 0.62 & 0.5 \\
 \hline
 $f_0$                          & 0.5   & 0.51  & 0.53 & 0.60 & 0.62 & 0.69 & 0.76 & 0.88 & 0.94 & 1 \\
 $f_{0.5}$                      & 0.87  & 0.88  & 0.77 & 0.76 & 0.75 & 0.75 & 0.75 & 0.74 & 0.74 & 0.71 \\
 $f_1$                          & 1.5   & 1.51  & 1.12 & 0.95 & 0.91 & 0.81 & 0.74 & 0.62 & 0.58 & 0.5 \\
 $f_{1.5}$                      & 2.6   & 2.61  & 1.63 & 1.19 & 1.10 & 0.87 & 0.73 & 0.52 & 0.46 & 0.35 \\
 $f_2$                          & 4.5   & 4.50  & 2.37 & 1.50 & 1.32 & 0.94 & 0.72 & 0.44 & 0.36 & 0.25 \\
 $f_3$                          & 13.5  & 13.38 & 5.02 & 2.37 & 1.92 & 1.09 & 0.69 & 0.31 & 0.22 & 0.125 \\
 \hline
 \end{tabular}
\end{table}

As it was spoken, the factor $k$ depends also on the spin $a_*$,
but this dependence is weak: $k = 1/3$ for $a_* = 0$ and $k= 1/2$
for $a_* = 1$. This dependence does not break the monotonic
increase of $B_{\parallel}(R_{\lambda})$ with the decrease the
spin. For this reason, further we shall neglect this weak effect.
Thus, the Faraday depolarization, which is proportional to
$B_{\parallel}(R_{\lambda})$, increases with the spin decrease.
Therefore, the observed degree of polarization $p_{obs}$ will
decrease.

As it is seen from the derivation of formulae (12) and (13), the
effect of polarization decreasing for slow rotating black holes as
compared with the fast rotating ones, mainly is determined by the
fact, that in the first case the orbit $r_{ms}$ is located far
from the disk center than for the second case. This effect became
more strong with grow of index $n$ (for $n=0.5, 1, 1.5, 2, 3$ the
ratio $f_{n}(0)/f_{n}(1)$ acquires, the values 1.22, 3, 7.34, 18,
108, respectively).

If one considers the objects with equal masses $M_{BH}$ and
luminosities $L_{bol}$, then it is necessary to take into account
that the parameter $\varepsilon$ also depends on the spin $a_*$.
In this case the value of magnetic field
$B_{\parallel}(R_{\lambda})$ (and also the polarization) is
determined by $\Phi_n(a_*) = f_n(a_*) \varepsilon^{n/3 -
1/2}(a_*)$ ($B_{\parallel}(R_{\lambda}) \sim \Phi_n$). The values
$\Phi(a_*)$ for $a_*=0$ and 1 are given in Table 2.

For most probable values of index $n\approx 1.5 - 2$, magnetic
field in the emission region for Schwarzschild black hole is
approximately to ten times higher than for fast rotating Kerr
hole. Clearly, in the first case  the Faraday depolarization is
considerably higher than in second case. This means that the
observed degree of polarization from active galactic nucleus (AGN)
with the Kerr black hole will be higher than that for
Schwarzschild's hole, if we compare the systems with approximately
equal masses and luminosities.

It is of interest to see what changes if to suppose that potential
law (8) of magnetic field decreasing takes place directly from the
black hole horizon, i.e. one exists the relation:

\begin{equation}
 B_{\parallel}(r)= B_H \left(\frac{R_H}{r}\right)^n =
 2^{-n}(1+\sqrt{1-a^2_*})^n B_H\left(\frac{r_g}{r}\right)^n.
 \label{eq14}
\end{equation}

\noindent Evidently, in this case we have to take $k=1$ and $q=1$.
The magnetic field dependence from $a_*$
$B_{\parallel}(R_{\lambda})$ exists due to factor
$\Psi_n(a_*)=(1+\sqrt{1-a^2_*})^{(n-1)}$ (this is analog to
$f_n(a_*)$).

In this case at $n=0.5, 1, 1.5, 2, 3$ the ratio
$\Psi_{n}(0)/\Psi_{n}(1)$ acquires the values  0.71, 1, 1.41, 2,
4. These values are considerably lower than those for
$f_{n}(0)/f_{n}(1)$. For ratio
$\Psi_{n}(a_*)\varepsilon^{n/3-1/2}$ (the analog $\Phi_{n}(a_*)$)
at $a_*=0$ to the value at $a_*=1$ we obtain $\approx 1.4$ for all
values of $n$. This is also is much higher than the corresponding
values $\Phi_{n}(0)/\Phi_{n}(1)$. Thus, the strong difference
between polarizations in Schwarzschild and Kerr cases really is
due to large difference between the radiuses of first stable
orbits in these cases.

The results of Table 1 show that there exists the dependence of
polarization degree grow with the grow of black hole spin. It
seems for ensemble of AGNs with approximately close values of
masses and luminosities we may to consider the spin $a_*\approx 1$
in systems with most high polarization, but with $p_{max}<
11.7\%$. For sufficiently large ensembles one has compare the
observing polarization with the $\mu$-averaged values of formula
(5). Such comparison might give the estimates for
$\Phi_{n}(a_*)=f_n(a_*)\varepsilon^{n/3-1/2}(a_*)$ for
intermediate between $a_*=0$ and $a_*=1$ values of spin, i.e. give
the estimates for parameter $\varepsilon(a_*)$, and also for
parameter $n$. It better for this to use the values of
polarization in different wavelengths. This difficult problem is
not considered in our paper.

\section{The spin dependence of polarization from the region of
 broad lines} 

In this case the characteristic distance $R_{BLR}$ of region
emitting broad lines is determined by the expression (see Shen
2009):

\begin{equation}
 R_{BLR} = 2.1 \times 10^{13}
 \left(\frac{M_{BH}}{M_{\odot}}\right)^{1/2}
 \left(\frac{L_{bol}}{L_{edd}}\right)^{1/2} =
 5.525 \times 10^4 \sqrt{\varepsilon} \sqrt{\dot{M}}.
 \label{eq15}
\end{equation}

\noindent Instead of formula (11) we obtain

\begin{equation}
 \frac{r_g}{R_{BLR}} = 10^{-7.85}
 \left(\frac{M_{BH}}{M_{\odot}}\right)^{1/2}
 \left(\frac{L_{bol}}{L_{edd}}\right)^{-1/2} =
 5.34 \left(\frac{M_{BH}}{M_{\odot}}\right)
 \frac{1}{\sqrt{\varepsilon} \sqrt{\dot{M}}}.
 \label{eq16}
\end{equation}

\noindent Substituting this expression into general formula (8),
we obtain the following expression for\\ $B_{\parallel}(R_{BLR})$:

\[
 B_{\parallel}(R_{BLR}) =
 \frac{k q^n \sqrt{k_m}}{1 + \sqrt{1 - a_*^2}} 10^{8.8 - 7.85 n}
 \left(\frac{M_{BH}}{M_{\odot}}\right)^{n/2 - 1/2}
 \left(\frac{L_{bol}}{L_{edd}}\right)^{-n/2 + 1/2}
 \frac{1}{\sqrt{\varepsilon}} =
\]
\begin{equation}
 = 10^{0.22 + 0.73n}
 \frac{k \sqrt{k_m}}{1 + \sqrt{1 - a_*^2}} q^n
 \left(\frac{M_{BH}}{M_{\odot}}\right)^{n - 1}
 \frac{(\dot{M})^{1/2 - n/2}}{(\sqrt{\varepsilon})^n}.
 \label{eq17}
\end{equation}

\noindent So, in the case of known luminosity $L_{bol}$ and mass
$M_{BH}$

\[
 B_{\parallel}(R_{BLR}) \sim
 \frac{k q^n \varepsilon^{-1/2}}{1 + \sqrt{1 - a_*^2}} =
 \bar{\Phi}_{n}(a_*)=f_{n}(a_*)\varepsilon^{-1/2},
\]
\begin{equation}
 \varepsilon^{-1/2} = 4.188\,\, for\,\, a_* = 0,\,\,\,
 \varepsilon^{-1/2} = 1.543\,\, for\,\, a_* = 1.
 \label{eq18}
\end{equation}

\begin{table}
 \caption[]{Values of $\Phi_{n}(a_*)$ and $\bar{\Phi}_{n}(a_*)$.}
 \label{tab2}
 \centering
 \begin{tabular}{|c|c|c|c|c|c|}
 \hline
 $n$                                          & 0.5  & 1    & 1.5   & 2     & 3 \\
 \hline
 $\Phi_{n}(0)$                                & 2.25 & 2.42 & 2.60  & 2.79  & 3.23 \\
 $\bar{\Phi}_{n}(0)$                          & 3.64 & 6.28 & 10.9  & 18.8  & 56.4 \\
 \hline
 $\Phi_{n}(1)$                                & 0.94 & 0.58 & 0.35  & 0.22  & 0.08 \\
 $\bar{\Phi}_{n}(1)$                          & 1.09 & 0.77 & 0.55  & 0.38  & 0.19 \\
 \hline
 \hline
 $\Phi_{n}(0) / \Phi_{n}(1)$                  & 2.38 & 4.18 & 7.34  & 12.9  & 39.8 \\
 $f_{n}(0)/f_{n}(1)$                          & 1.22 & 3    & 7.34  & 18    & 108 \\
 $\bar{\Phi}_{n}(0) / \bar{\Phi}_{n}(1)$      & 3.34 & 8.16 & 20    & 50    & 297 \\
 \hline
 \end{tabular}
\end{table}

\noindent Remember that for emission region in continuum the value
$\varepsilon^{-1/2}$ is replaced by $\varepsilon^{n/3 - 1/2}$. In
Table 2 we presented  $\Phi_{n}(a_*)$ and $\bar{\Phi}_{n}(a_*)$
for $a_* = 0$ and $a_* = 1$. It is seen that magnetic field in
broad line emission region is also considerably lower for Kerr
holes than that for Schwarzschild hole. We stress once more, that
this takes place mainly due to that radius $r_{ms}$ for Kerr black
holes is located more far from the emission region, and
consequently the magnetic field is lower than in the case of
non-rotating Schwarzschild holes with its $r_{ms} = 3r_g$.

\section{Conclusion} 

Using the model that the linear polarization of radiation from AGN
is due to multiple scattering in magnetized accretion disks, we
demonstrate that the polarization strongly depends on the spin of
central black hole. For fast rotating Kerr-type black holes the
linear polarization is considerably higher than that for
non-rotating Schwarzschild's holes. This is mainly occur due to
fact that the radius of first stable orbit  for Kerr holes is more
far from the emission region $R_{\lambda}$ than that for
Schwarzschild holes with This result is based on the supposition (
seems very obvious) that the power law of magnetic field decrease
takes place from the first stable orbit $r_{ms}$, which is greater
($r_{ms}=3r_g$) for Schwarzschild hole. The far located the
emission region from the first stable orbit the less is magnetic
field in this region, and less the depolarization due to Faraday
rotation effect. It is known that bolometric luminosity of
accretion disk near fast rotating Kerr black hole is considerably
higher than that for non-rotating hole with the same mass. It
means that bright AGNs have to demonstrate higher polarization of
optical radiation than weak AGNs.

\section*{Acknowledgements}

This research was supported by the program of Prezidium of RAS
''Origin and Evolution of Stars and Galaxies'', the program of the
Department of Physical Sciences of RAS ''Extended Objects in the
Universe'', by the Federal Target Program ''Scientific and
scientific-pedagogical personnel of innovative Russia'' 2009-2013
(GK 02.740.11.0246) and by the grant from President of the Russian
Federation "The Basic Scientific Schools" (NSh-3645.2010.2).

\end{document}